\documentclass[%
reprint,
 amsmath,amssymb,
 aps,nofootinbib,   
]{revtex4-1}
\usepackage{bm}
\PassOptionsToPackage{linktocpage}{hyperref}
\usepackage[hyperindex,breaklinks]{hyperref}
\usepackage{enumitem}
\usepackage{slashed}

\renewcommand{\theta}{\vartheta}

\newcommand{\arXiv}[2]{\href{http://arxiv.org/pdf/#1}{{\tt [#2/#1]}}}
\newcommand{\arXivold}[1]{\href{http://arxiv.org/pdf/#1}{{\tt [#1]}}}

\usepackage{array}
\usepackage{mathtools}

\usepackage{etoolbox}

\begin{document} 

\title{Inflation and Decoupling}

\author{Gia Dvali$^{a,b}$, Alex Kehagias$^{c}$,  Antonio Riotto$^{d,e}$
} 
\affiliation{
	$^a$Arnold Sommerfeld Center, Ludwig-Maximilians-Universit\"at, Theresienstra{\ss}e 37, 80333 M\"unchen, Germany
}
\affiliation{
	$^b$Max-Planck-Institut f\"ur Physik, F\"ohringer Ring 6, 80805 M\"unchen, Germany
}
\affiliation{ 
	$^c$  Physics Division, National Technical University of Athens 15780 Zografou Campus, Athens, Greece }

\affiliation{$^d$  D\'epartement de Physique Th\'eorique and Centre for Astroparticle Physics (CAP), \\
Universit\'e de Gen\`eve, 24 quai E. Ansermet, CH-1211 Geneva, Switzerland }
\affiliation{$^e$ INFN, Sezione di Roma, Piazzale Aldo Moro 2, 00185, Roma, Italy}


\begin{abstract} 
\noindent
Decoupling of heavy modes in effective low energy theory 
is one of the most fundamental concepts in 
physics. It tells us that modes 
must have a negligible effect on the physics of gravitational backgrounds with curvature radius 
larger than their wavelengths. Despite this, there exist  
claims that trans-Planckian modes put severe bound on 
the duration of inflation even when the Hubble parameter 
is negligible as compared to the Planck mass. If true, this would mean that 
inflation violates the principle of decoupling or  at least requires its reformulation. 
 We clarify the fundamental misconception on which these
bounds are based and respectively refute them. Our conclusion is that 
inflation fully falls within the validity of a reliable effective field theory treatment 
and does not suffer from any spurious trans-Planckian problem.    
 \end{abstract}

\maketitle
\noindent
One of the most fundamental concepts in physics tells us that modes
operating at some $UV$-wavelength 
$L_{UV} \equiv \Lambda_{UV}^{-1}$ decouple from the low energy processes taking place at $IR$-scale 
$L_{IR} \gg L_{UV}$. That is, in an Effective Field Theory 
(EFT), formulated at a low energy scale $L_{IR}$, the short wavelength physics sums itself up in effective coefficients of renormalizable interactions and in infinite 
series of non-renormalizable contact interactions suppressed by 
powers of $L_{UV}$.  At the same time, any long-distance 
correlator generated by the exchange of $UV$-modes, is exponentially 
suppressed by a factor of the type ${\rm exp}(-L_{IR}/L_{UV})$. 
 Although this concept is usually referred to as {\it Wilsonian 
 decoupling}, it goes beyond the regime of  standard Wilsonian $UV$-completion.
 For example, it remains equally correct also if $UV$-completion 
 above the cutoff $\Lambda_{UV}$
happens via {\it classicalization} phenomenon \cite{class, self}. 
In such theories the states of energy $M \gg \Lambda_{UV}$, instead of being represented by 
 single-particle states of wavelength  $L \ll L_{UV}$, are  
 described by classical configurations
  composed of many soft quanta
  of wavelength 
 $L  \gg L_{UV}$ \cite{classN, NP, 2N}.  
 That is, putting it shortly, Wilsonian decoupling holds even when the 
 $UV$-completion is non-Wilsonian. \\

 In the first part of the present paper, we shall stay within the range  of validity of  the standard Wilsonian picture of $UV$-completion, whereas  
 in the second part we shall extend our results to a non-Wilsonian domain.   \\
 
  A classic example of the decoupling at work is provided by Einstein's theory of gravity. Indeed, already 
  in lowest order, the interactions of graviton are
  represented by non-renormalizable operators suppressed by 
  powers of the Planck mass $M_P$. Despite this fact, 
  the effective low energy theory allows to predict phenomena 
  with extraordinary accuracy, without any need to worry about 
  the quantum gravitational corrections from the Planck scale physics. 
  The reason is that the quantum gravitational coupling 
  among the elementary particles of wavelength $L \gg L_P$ is 
  extremely small, 
 \begin{equation} \label{alpha}
   \alpha_{GR} = \frac{1}{(LM_P)^2} \,. 
   \end{equation} 
 Correspondingly, for many sources,  the expansion in series of $\alpha_{GR}$ is highly reliable.  
  For instance, for a graviton of wavelength comparable to earth-moon distance ($L \sim 10^{10}$cm), the coupling is 
 $\alpha_{GR} \sim 10^{-86}$. Obviously, only the very heavy classical sources (such as the earth) can compensate this enormous suppression. 
 The bonus is that for such sources the effect 
 (e.g., the lunar orbit) can be computed  extremely accurately. 
 Obviously, the dominant contribution into the exchange comes 
 from the gravitons of the wavelengths $\sim L$.   
 At the same time, the corrections from gravitons of the shorter 
wavelengths is exponentially small.  \\

In general, the criterion of insensitivity towards the short-scale physics can be formulated in terms of the curvature invariants. Namely, for the classical backgrounds 
of large curvature rarius $L \gg L_P$, the corrections from Planck-scale 
physics is expected to be highly suppressed.  \\
 
A well-known manifestation of the above concept is the success of the inflationary paradigm \cite{Guth,lrr}. According to Guth's original idea,  
our Universe  underwent through a De Sitter like epoch
during which the scale factor $a(t)$ was increasing exponentially 
$a(t) \propto {\rm e}^{Ht}$. Here $H$ is the inflationary Hubble parameter 
which sets the curvature radius $H^{-1}$ and is approximately constant in cosmic time $t$.  In this way, the inflation addresses the most fundamental 
cosmological questions such as the horizon and flatness. \\

Obviously, the decoupling principle tells us that 
as long as $H \ll M_P$, the inflation can be treated
reliably within the low energy EFT.
The leading corrections from Planck wavelength physics may 
come in form of the higher curvature invariants. 
These, however, are suppressed by powers of 
$H^2/M_P^2$ relative to Einstein, 
see e.g. \cite{ex}.
 \\

Now, it is very important to understand that  the decoupling principle 
fully permits 
the microscopic physics to have significant 
macroscopic effects on sufficiently long time-scales. 
For example, a macroscopic tank of water can be emptied 
due to proton-decay mediated by
$UV$-physics. 
 Likewise, the time-scale on which the microscopic quantum corrections 
 to De Sitter and inflation become important is given 
 by the following {\it quantum break-time} \cite{Ncoh, QB},
 \begin{equation} \label{macro} 
  t  \sim H^{-1} \frac{M_P^2}{H^2} \, .  
\end{equation}  
After this time, in general, the microscopic 
theory must be taken into account.  
 Without entering into much details, here, we can justify (\ref{macro})
 by the following simple short-cut dimensional argument. 
  Indeed, the largest rate of a quantum process taking place 
 at energy $H$ and controlled by a cutoff scale $\Lambda_{UV} = M_P$ is given by 
  $\Gamma \sim  H^3/M_P^2$.  Then, it is obvious that (\ref{macro}) 
  is a minimal time-scale, $t = \Gamma^{-1}$, required for such a 
  process to become effective.  Of course, in theories with a lower cutoff $\Lambda_{UV}$ 
 the time scale is shortened accordingly. \\

For the future comparison, it is important that microscopic considerations lead \cite{Ncoh} 
 to derivation of a second, logarithmic, time-scale,
 \begin{equation} \label{Nlog}
  t \sim H^{-1} \ln M_P^2/H^2 \,.   
 \end{equation} 
This was obtained as a characteristic  time after which the 
De Sitter quantum state can become one-particle entangled\footnote{
The scale (\ref{Nlog}) was proposed in \cite{Ncoh} as the De Sitter analog of so-called information scrambling limit \cite{HPr}. The quantum field theoretic 
meaning given to it in \cite{Scr} is of a lower limit on time after which a 
system with Lyapunov instability starts developing chaos.
These interpretations are secondary for the present discussion
and we shall not enter there.  An interested reader is referred to the original papers cited above and references therein.}. The main thing for the present discussion is that after 
(\ref{Nlog}) the back reaction to De Sitter is still negligible and cannot affect 
the validity of inflation. \\ 

  To summarize, in accordance with the decoupling principle, 
  the microscopic effects from quantum gravity do not invalidate 
  the EFT treatment of De Sitter on the time-scales shorter than (\ref{macro}). This  gives  a large domain of validity for standard 
inflation.  For example, for $H \sim 10^{13}$~GeV,  the available number of 
e-foldings, ${\mathcal N}_e\equiv tH$,
would be over $ {\mathcal N}_e \sim 10^{12}$.
\\

 Despite the above, in the literature one encounters 
 discussions that can be referred to as  the {\it trans-Planckian 
 problem of inflation}, see e.g.  \cite{TP,kin,TPrev}.  The argument can be summarized as follows. 
 Let us consider an inflationary fluctuation that is detected at some later 
 time, e.g., today. 
  At the moment of crossing 
 outside the inflationary Hubble patch this fluctuation had a wavelength
 $L \sim H$. 
  This wavelength is a result of the stretching due to the exponential 
 expansion of the scale factor.  
 Thus,  scaled back in time by ${\mathcal N}_e$ inflationary $e$-foldings, 
 it shrinks to a size $L_{in} = L e^{-{\mathcal N}_e}$.  Hence, 
 if inflation lasted longer than 
 \begin{equation} \label{Nmax}
  {\mathcal N}_e = \ln M_P/H, 
 \end{equation}  
  some perturbations 
would inevitably reach the trans-Planckian wave-lengths 
in the past.  
This reasoning prompted the authors of Ref. \cite{TCC}  to conjecture that
(\ref{Nmax}) must be accepted as an upper bound on the number of inflationary e-foldings. This is the so-called
{\it Trans-Planckian Censorship Conjecture}.  \\

The coincidence of the time-scale (\ref{Nmax}) with 
(\ref{Nlog}) is obvious but the interpretation
given to it by Ref. \cite{TCC} is fundamentally different as it signals a complete breakdown of inflation as EFT.     
If this were true, it would imply that inflation violates the decoupling 
principle, or the least, demands its fundamental rethinking and  reformulation.  
Indeed, according to this view, for a low energy observer 
operating at distance $H^{-1}$, it takes only a logarithmic time 
to be strongly affected by the Planck scale physics!  
 This would be a truly remarkable prospect. \\
 
  Unfortunately, the above is not the case and neither (\ref{Nlog}) 
 nor (\ref{Nmax}) 
  represent the sensible bound on the duration of inflation. 
  The reason lies in a certain misconception slipped though   
 the above thought experiment during which the inflationary perturbations
 have been scaled back in time. \\

   The point is that it makes no physical sense 
to scale a given fluctuation arbitrarily far back in time. 
This is because, prior to a certain initial moment, the fluctuation 
simply {\it did not  exist}. 
That is, an overwhelming majority of De Sitter quantum fluctuations 
are produced with wavelengths $L \sim H^{-1}$. 
Only an exponentially-suppressed fraction $\sim e^{-1/(HL)}$  
is created with wavelengths $L \ll H^{-1}$.  Obviously, it is misleading 
to scale the modes back in time past their ``birth date".   Or putting it 
differently, once we shrink the mode beyond  $H^{-1}$, we must weight 
it with a probability that the mode was already {\it real}  and not
the part of the {\it unmaterialized vacuum spectrum}. 
This probability is suppressed by the above exponential factor even much {\it earlier} the mode shrinks down to the Planck length. 
This suppression was not taken into account in the reasoning 
that leads to the bound (\ref{Nmax}).  
\\ 

 In order to grasp the above more clearly, let us reduce to bare essentials the 
 physical mechanism of creation of De Sitter quantum fluctuations. Let us consider a 
 quantum field (e.g. a graviton)  in the De Sitter Universe. An each momentum mode of the field represents a quantum oscillator. The main effect of the 
 expansion is that the frequency of the oscillator  $\omega$  changes in time. This change leads to a particle-creation since the modes that 
 at certain moment of time $t$ are in the vacuum, at the later times,  
are above it.  This is a stationary process and the only control-parameter is 
the rate by which the frequency $\omega$ changes in time. 
 This rate is set by the Hubble $H$ and therefore is constant.  As a result,  the modes that are 
 created out of the vacuum have frequencies $\omega \sim H$. The 
 modes of a higher frequency 
 $\omega \gg H$ are exponentially suppressed. No matter how long an observer shall wait,  the production of the high frequency modes shall not 
  become more probable. In particular,  if $H \ll M_P$, an observer
  shall see the production of Planck frequency modes 
  extremely rarely: One per Hubble patch per exponentially long time 
  $\sim H^{-1} e^{-1/(HL)}$.  \\

  This way of looking at things makes it 
  obvious that there cannot possibly be any trans-Planckian problem 
  in inflation. The reason is simple: a soft inflationary background cannot produce the Planck 
  energy quanta. For comparison,  an inflationary Universe with 
  Hubble $H \sim 10^{13}$ GeV is less sensitive to the Planck scale 
  physics than the energy levels of the Hydrogen 
 atom are to the weak interaction!   \\
 
 In order to make the point sharper, let us monitor the de Sitter quantum fluctuation by a parallel particle physics process.  
 For definiteness, we take our current Universe which is known to be 
 De Sitter like. In this Universe we shall consider the two parallel quantum processes. 
 One is the usual process of creation of particles (gravitons) in De Sitter 
 that was already considered above. 
  Another one, is a  decay process 
 mediated by a high dimensional operator.  
 For instance, let us take a hypothetical proton-decay in which one of the final state particles is a photon. The two processes are analogous 
 in the following sense. Both interactions are suppressed by 
 respective cutoffs and both suppressions are compensated by 
 the magnitude of the respective macroscopic sources. In case of gravity, this is the energy of the entire De Sitter patch. In case of the proton-decay, 
 the source is a large tank 
 of water that contains many protons. \\

  Now,  after being produced in an expanding De Sitter universe, both 
 the graviton and the photon will get redshifted and after some time can be detected by a future observer  (Alice).   Now, if Alice will
scale the two modes back to an indefinite past, 
 she will arrive to a wrong conclusion that in some distant past both modes had the trans-Planckian wavelengths. For the photon this is obviously wrong since it was 
 produced in a decay of a proton. Then, obviously the same must be 
 true about the graviton since the two particles were produced simultaneously. \\
 
 This completes the first part of our discussion in which we
 clarify the source of fictional non-decoupling problem arising by a naive 
 past-scaling of 
 modes. As explained, this scaling does not take into account  the suppressed probability for materializing perturbations with the wavelengths shorter than Hubble.  \\

 We now wish to clarify the second grave misconception that arises 
 by the above naive past-scaling of modes to so-called 
{\it trans-Planckian} regime. 
The trans-Planckian regime is usually understood as the past epoch in 
which a wavelength of a given mode $L$ was shorter than the 
Planck length.  Such a definition already carries in it a dangerous  
fuzziness as we shall now explain.   
In ordinary renomalizable theories with Wilsonian $UV$-completion (e.g., 
such as QCD), one can in principle probe an arbitrarily short distance
scale $L$. 
All is needed, is to arrange a $2\rightarrow 2$ particle scattering with
momentum-transfer $\sim L^{-1}$. That is, in such theories we can localize 
an elementary particle within an {\it arbitrarily small} region of space $L$ 
provided we invest energy $M \sim 1/L$. \\ 
 
In contrast, in  Einstein gravity such a reasoning works only till the Planck length
$L_P$. That is, in Einstein gravity an excitation 
of a center of mass energy $M \gg M_P$ cannot be described as a  single-particle state of any elementary quantum field \cite{self}. 
Instead, it {\it classicalizes}  and represents a black hole. 
Correspondingly, the minimal localization radius is set by the 
{\it classical}  gravitational radius $\sim ML_P^2 \gg L_P$. The latter 
exceeds 
 the quantum Compton wavelength $1/M$. This fact already signals that the object is classical.  This phenomenon is a fundamental property of Einstein gravity and is completely independent of the details of trans-Planckian 
 theory \cite{self}.  It tells us that, no matter how profound the 
 $UV$-theory is, the 
 heavy modes must decouple since from the low energy perspective they represent classical black holes. \\
 
 Once again, in order to avoid misunderstanding, we stress that this 
 discussion is {\it not}  about advocating any particular scenario of 
 $UV$-completion. Rather, we wish to make it clear that classicalization of heavy modes into black holes - which is not an assumption but a property 
 of Einstein theory - 
 ensures that the decoupling holds {\it universally}  regardless
of the properties of $UV$-theory.  
 To put it differently, irrespectively what 
miraculous properties one mentally assigns to $UV$-theory, 
 the decoupling of heavy modes  cannot be questioned. 
 The fact that such modes represent black holes, is fully controlled by $IR$-theory.  
  \\  
 
 In order to remove any doubts whether the black holes of $IR$-theory 
 can be eliminated by  assuming some exotic modifications of dispersion relations 
 at trans-Planckian  distances,  we wish to point out that such modifications
 are impossible without sacrificing the most fundamental consistency properties such as causality and positivity of norm and energy.  
As explained in details in \cite{self}, the basic analyticity properties 
severely restrict the pole structure of any possible modification 
of graviton propagator. In particular, any $UV$-modification of the graviton dispersion relation that would abolish classicalization of
high energy states into $IR$-black holes, requires appearance of ghost 
poles in graviton propagator and therefore is excluded. 
This restriction is non-perturbative and follows from the properties of 
spectral representation of most general graviton propagator.  
We shall naturally not be interested in such inconsistent modifications.
Thus, again, regardless what mechanism is responsible for  $UV$-completion above the energy $M_P$, the excitations with such center of mass energies  
classicalize into black holes.  This information suffices for our 
further analysis.
\\

Now,  it is clear that when one talks about scaling the inflationary perturbations back in time towards the trans-Planckian regime, in reality, one talks about  scaling them past the point 
of their {\it classicalization}.  
  The only way one could give a consistent physical meaning to such a scaling is to weight it by a probability of materializing the De Sitter perturbation of trans-Planckian energy outside of their gravitational radius.    
  By now, it should be obvious for a reader that by any sensible estimate 
  this probability must be totally negligible. \\
  
  For illustrative purposes, we shall estimate it for a limiting case 
  when the energy of a would-be trans-Planckian mode is comparable  to the energy of the entire Hubble patch, $M  \sim M_P^2/H$. 
  The gravitational radius of a corresponding black hole is 
  obviously of the order of the Hubble radius $H^{-1}$. 
  We shall now estimate the probability of producing such a mode 
  in De Sitter in its particle form. That is, the probability that De Sitter materializes modes outside of the gravitational radius of 
 center of mass energy $M$. 
 We shall perform the estimate in three different ways and show that they all agree. \\
  
  The first way is to think of De Sitter as a (approximate) thermal bath with Gibbons-Hawking temperature  $T_{GH} \sim H$ \cite{GH}. 
  The probability of producing a mode 
 of energy $M$  is then exponentially suppressed by a Boltzmann factor, 
  \begin{equation} \label{boltzmann}
    \Gamma \sim  e^{-M/T_{GH}} \sim  e^{-M_P^2/H^2}\,.
   \end{equation} \\
   The second method is to think of the entropy-suppression.  
  Indeed, by giving away a half of its energy into a single elementary quantum of a very low entropy, the total entropy of the system decreases. 
 Namely, the De Sitter Gibbons-Hawking entropy,  $S_{GH} =  M_P^2/H^2$, decreases by an order-one  fraction.  This decrease must result into an entropy suppression price of the process,  $\sim e^{-S_{GH}}$, which fully matches 
  (\ref{boltzmann}). \\
  
  Finally, perhaps the most systematic estimate of the transition
  is  within the picture in which the De Sitter Hubble patch is resolved as a coherent state $|{\rm dS}\rangle$ of soft gravitons of frequency 
  $H$ and occupation number $N \sim  M_P^2/H^2$
  \cite{Ncoh,QB}.
  Following these works, the transition can be computed as an $S$-matrix process 
  in which  order $N$ soft gravitons merge into a single (or a pair) 
  of very hard gravitons of energy $M$. Using an explicit computation  
 of the multi-particle graviton amplitude performed in \cite{2N}, the rate of the process was obtained 
 in \cite{QB} and it is given by 
 \begin{eqnarray}
\Gamma\sim N!\,\alpha_{GR}^N\sim e^{-N} \label{lN}
 \end{eqnarray}
 in the large $N$.
  This again matches  (\ref{boltzmann}).  In other words,  the classicalization forces the trans-Planckian mode to ``dive'' into the vacuum whereas  the probability to survive in any other form is  exponentially small. Notice, in this language the suppression 
has the following physical meaning: 
 it represents an exponential suppression characteristic of a quantum transition between many soft and few hard quanta.  \\

 We thus observe that the three different estimates of the transition 
 between a 
 would-existing trans-Planckian graviton and De Sitter (Boltzmann, entropy and $S$-matrix process) give one and the same exponential 
  suppression (\ref{boltzmann}). 
    For $H \ll M_P$ this is vanishing 
  beyond any repair.  \\

 In conclusion, the inflationary Universe is subject to the same laws of decoupling as any other physical system within the validity of EFT treatment.     
  In particular, there is no trans-Planckian problem in inflation and 
  the bound (\ref{Nmax}) is spurious.  
 Instead, the Wilsonian decoupling indicates that 
 the time-scale of validity of De Sitter  is   
 not shorter than (\ref{macro}) which gives a significant room for 
 inflation. 
 \\

  \vskip 0.2cm
 \noindent
\centerline{\it Acknowledgements}
\vskip 0.2cm
\noindent
We thank R. Brandenberger and W. Kinney  for discussions on the Trans-Planckian Censorship Conjecture. 
 Also,  special thanks are due to Goran Senjanovic. 
 The work of G.D. was supported in part by the Humboldt Foundation under Humboldt Professorship Award, by the Deutsche Forschungsgemeinschaft (DFG, German Research Foundation) under Germany's Excellence Strategy - EXC-2111 - 390814868,
and Germany's Excellence Strategy  under Excellence Cluster Origins.
A.R. is  supported by the Swiss National Science Foundation (SNSF), project {\sl The non-Gaussian Universe and Cosmological Symmetries}, project number: 200020-178787.


\begin{thebibliography}{10}

 \bibitem{class}  
  G.~Dvali, G.~F.~Giudice, C.~Gomez and A.~Kehagias,
  JHEP {\bf 1108}, 108 (2011),
  \arXiv{1010.1415}{hep-ph}. 
  
  \bibitem{self}
G.~Dvali and C.~Gomez,
\arXiv{1005.3497}{hep-th}.  

    
  \bibitem{classN}  
  G.~Dvali, C.~Gomez and A.~Kehagias,
  JHEP {\bf 1111}, 070 (2011)
  \arXiv{1103.5963}{hep-th}.
  
 \bibitem{NP} 
  G.~Dvali and C.~Gomez,
  Fortsch.\ Phys.\  {\bf 61}, 742 (2013)
  \arXiv{1112.3359}{hep-th}.  
  

 \bibitem{2N} 
  G.~Dvali, C.~Gomez, R.~S.~Isermann, D.~L\"ust and S.~Stieberger,
    Nucl.\ Phys.\ B {\bf 893}, 187 (2015)
  \arXiv{1409.7405}{hep-th}.


 
 
 \bibitem{Guth} A.~Guth, Phys. Rev. D23 (1981)
347.

\bibitem{lrr} D.~H.~Lyth and A.~Riotto,
  Phys.\ Rept.\  {\bf 314}, 1 (1999)
  \arXivold{hep-ph/9807278}.

 \bibitem{ex}  N.~Kaloper, M.~Kleban, A.~E.~Lawrence and S.~Shenker,
  Phys.\ Rev.\ D {\bf 66}, 123510 (2002)
  \arXivold{hep-th/0201158}.
  
  \bibitem{Ncoh}
G.~Dvali, C.~Gomez,
  JCAP 1401 (2014) 023, 
\arXiv{1312.4795}{hep-th};

\bibitem{QB} G.~Dvali, C.~Gomez, S.~Zell,	
  JCAP 1706 (2017) 028, 
  \arXiv{1701.08776}{hep-th}. 

\bibitem{HPr} P.~Hayden and J.~Preskill, 
JHEP 0709 (2007) 120,
 \arXiv{0708.4025}{hep-th}.
  
\bibitem{Scr}
G.~Dvali, D.~Flassig, C.~Gomez, A.~Pritzel, N.~
Wintergerst,  Phys. Rev. D \text{88} (2013) 12, 124041,
  \arXiv{1307.3458}{hep-th}.  
  
  
  

\bibitem{TP} J.~Martin and R.~H.~Brandenberger,
   Phys.\ Rev.\ D {\bf 63}, 123501 (2001)
  \arXivold{hep-th/0005209}.
  
  \bibitem{kin} R.~Easther, B.~R.~Greene, W.~H.~Kinney and G.~Shiu,
Phys. Rev. D \textbf{66}, 023518 (2002)
\arXivold{hep-th/0204129}.

\bibitem{TPrev} R.~H.~Brandenberger and J.~Martin,
Class. Quant. Grav. \textbf{30}, 113001 (2013)
\arXiv{1211.6753}{astro-ph.CO}.

   \bibitem{TCC} 
  A.~Bedroya, R.~Brandenberger, M.~Loverde and C.~Vafa,
  Phys. Rev. D \textbf{101} (2020) no.10, 103502
  \arXiv{1909.11106}{hep-th}.

  


  
  
  
 


\bibitem{GH}  G.W.~Gibbons, S.W.~Hawking, Phys.
Rev. D15 (1977) 2738.


%



\end{thebibliography}
\end{document}